\documentclass{aa}
\usepackage{graphicx}
\usepackage{txfonts}
\begin{document}
\title{The AMIGA sample of isolated galaxies}
\subtitle{ VI. Radio continuum properties of isolated galaxies:
a very radio quiet sample}
\author{S. Leon\inst{1,2}
\and
 L. Verdes-Montenegro\inst{2}
\and
J. Sabater\inst{2}
\and
D. Espada\inst{3}
\and
U. Lisenfeld \inst{4}
\and
A. Ballu\inst{5}
\and
J. Sulentic\inst{6}
\and
S. Verley\inst{7}
\and
G.~Bergond\inst{2}
\and
E. Garc\'{\i}a\inst{2}}

\institute{
Instituto de Radioastronom\'{i}a Milim\'etrica (IRAM), Avenida Divina Pastora 7, N\'{u}cleo Central, 18012 Granada, Spain\\
\email{leon@iram.es}
\and
Instituto de Astrof\'{\i}sica de Andaluc\'{\i}a, CSIC,
Apdo. 3004, 18080
Granada, Spain
\and
Institute of Astronomy and Astrophysics, Academia Sinica,  No.1, Roosevelt Rd, Sec. 4, Taipei 10617, Taiwan
\and
Departamento de F\'{\i}sica Te\'orica y del Cosmos, Facultad de Ciencias, Universidad de Granada, Spain
\and
\'Ecole Nationale Sup\'erieure de Physique, Universit\'e Louis Pasteur, Strasbourg, France
\and
Department of Astronomy, University of Alabama, Tuscaloosa, USA
\and
INAF-Osservatorio Astrofisico di Arcetri, Largo E. Fermi 5, 50125 Firenze, Italy
}
\date{Received XX; accepted XX}

\abstract
{
This paper is part of a series describing the results of  the AMIGA 
(Analysis of the interstellar Medium of Isolated GAlaxies) project, studying  the
largest sample of very  isolated galaxies in the local Universe.
}
{
The study of the radio properties of  the AMIGA sample is intended to characterize
the  radio continuum emission for a sample least
affected by local environment, thus providing a reference against
which less isolated and interacting samples can be compared.
}
{Radio continuum data at 
325, 1420 and 4850 MHz were extracted from the WENSS, NVSS/FIRST and GB6
surveys, respectively. The source extractions have been obtained from reprocessing  the
 data  and  new detections have been added to the
cross-matched detections with the  respective survey catalogs. We focus on the complete
AMIGA subsample composed of 719 galaxies.
}
{
 From the above four surveys a catalog of radio fluxes was obtained.
Comparison between the NVSS and FIRST detections indicates that the 
radio continuum is coming from
disk-dominated emission in spiral galaxies, in contrast to the results 
found in high-density environments where nuclear activity is more frequent.
The comparison of the radio continuum power with a comparable sample, 
which is however not selected with respect to its environment,
the Condon et al. UGC-SF sample of star-forming field galaxies, 
shows a lower mean value for the AMIGA sample.
We have obtained radio-to-optical flux ratios ($R$) using the NVSS radio 
continuum flux. The distribution of $R$
for the AMIGA galaxies is consistent with a sample dominated by
radio emission from star formation (SF) and a small number of 
Active Galactic Nuclei (AGN),
 with less than 3\% of the sample with $R > 100$.
We  derived the radio luminosity function (RLF)
and total power
density of the radio continuum emission for the AMIGA sample at 1.4 GHz, and compared them with 
results from other low redshift studies.
The Schechter fit of the RLF indicates a  major weight  of the low-luminosity galaxies.
}
{
The results indicate the  very low level of radio continuum emission in our sample of
isolated galaxies, which is dominated by mild disk SF.  It confirms hence
the AMIGA sample as a suitable template to
effectively quantify  the role of interactions in  samples extracted
from denser environments.}


\titlerunning{The AMIGA sample of isolated galaxies. VI}
\maketitle

\section{Introduction}

Although it is widely accepted that galaxy interactions stimulate
secular evolutionary effects in galaxies (e.g. excess star formation,
active nuclei, morphological/kinematic anomalies) the frequency/amplitude 
of the effects and processes driving them are not well quantified.
Prior to about 1970 there was a general belief that interactions did
not stimulate unusual activity in galaxies. Radio continuum surveys of 
galaxy pairs emission statistics opened a debate about the role of 
interactions in stimulating activity (Tovmassian \cite{tovmassian68}). The period around 
1970 saw several studies reporting a lack of radio continuum enhancement 
in interacting galaxies (Allen et al. \cite{allen73}; Wright \cite{wright74}) closely
followed by several surveys that provided evidence for an enhanced level
of emission from pairs (Sulentic \cite{sulentic76a}, \cite{sulentic76b}; Stocke \cite{stocke78}; Stocke et al. \cite{stockeetal78}). As
the fulcrum shifted in favor of interaction induced activity, the
debate shifted to the relative response of disk versus nucleus as
the source of enhanced emission (Hummel \cite{hummel80}, \cite{hummel81}). Eventually
evidence emerged for excess emission from both nuclei and disks
(Condon et al. \cite{condon82}). Not surprisingly spiral galaxies have
provided the strongest evidence for a radio enhancement, but recently,
evidence has also arisen for radio continuum enhancements in
early-type members of mixed  pairs (E/S0+S), possibly via
cross-fueling from a gas rich neighbor (Domingue et al. \cite{domingue05}).
Interactions apparently do not always lead to enhancement, as
evidenced by compact groups where radio emission is weaker than
average, although more strongly concentrated than in more normal galaxies
(Menon \cite{menon92}, \cite{menon95}, \cite{menon99}).

A better quantification of interaction enhancement as a function of
many different observables is needed before models can be properly 
constrained and refined. This requires a proper definition of 
non-interacting galaxy in order to determine the ``zero-point''
or levels of self stimulated emission (nature vs. nurture). 
The AMIGA project (Analysis of the interstellar Medium of Isolated 
GAlaxies) involves identification and
parameterization of a statistically significant sample of the most
isolated galaxies in the local Universe. It is a refinement of the
Catalog of Isolated Galaxies (CIG: Karachentseva \cite{karachentseva73}) which is
composed of 1050 galaxies located in the Northern hemisphere. A
major AMIGA goal involves characterization of different phases of the
interstellar medium in galaxies least affected by their
environment. 

So  far we have: 1) revised all CIG positions
(Leon \& Verdes-Montenegro \cite{leon03}); 2) summarized the redshift and
luminosity properties of the sample including derivation of an
improved Optical Luminosity Function (Verdes-Montenegro et al. \cite{verdes05}); 
3) provided POSS2 based morphologies for the sample including identification 
of remaining certain (32) and suspected (161)  interacting galaxies 
(Sulentic et al. \cite{sulentic06}); 4) reevaluated the IRAS mid-infrared and far-infrared properties 
of the sample (Lisenfeld et al. \cite{lisenfeld07}); and 5) performed a reevaluation 
and quantification of the degree of isolation for the sample (Verley 
et al. \cite{verley07a}, \cite{verley07b}). The revised CIG-AMIGA sample is reasonably complete 
($\sim$80\%) down to $m_{B{\rm -corr}} \sim 15.0$ (Verdes-Montenegro et al. 
\cite{verdes05}) and is currently the largest sample of isolated galaxies in the local
Universe. These are galaxies whose structure and evolution have been
driven largely or entirely by internal rather than external forces
(i.e. as close to pure nature as exists in a local Universe where
nurture plays many roles). The data are being released and 
periodically updated at {\tt http://www.iaa.es/AMIGA.html}
 where a VO (Virtual Observatory) interface
with different query modes has been implemented.

We report here on the radio continuum properties of the AMIGA sample 
with presentation organized as follows: Section~\ref{sectData} describes
the radio surveys we used along with the (reprocessed) data and the
final catalogs. Section~\ref{sectRadioC} studies the radio continuum
characteristics of the AMIGA sample including  the correlation
between radio and optical luminosities, spectral index, as well as
comparison between NVSS and FIRST measures. The Radio Luminosity 
Function (RLF) and the power
density are presented and compared with other galaxy samples in
Sect.~\ref{sectLF}. Section~\ref{sectDisc} summarizes the main
results.

\section{The radio data\label{sectData}}

Several radio surveys now cover a large fraction of the sky with
detection limits of a few mJy and  spatial resolutions from a few
arcseconds to a few arcminutes. Four Northern sky surveys are 
available: 1) the Westerbork Northern Sky Survey (WENSS) at 
325~MHz; 2) the NRAO VLA Sky Survey (NVSS) at 1.4 GHz; 3) the Faint 
Images of the Radio Sky at Twenty-cm survey (FIRST)  at 1.4 GHz; and 
4) the Green Bank surveys (GB6) at 4.8 GHz. We cross-correlated the 
optically improved AMIGA positions (Leon \& Verdes-Montenegro \cite{leon03}) 
with  radio source positions from the above catalogs and further 
increased detection fractions by reprocessing source extractions from 
the calibrated data. Extractions were performed using the dedicated software
SExtractor (Bertin \& Arnouts \cite{bertin96}) with a threshold of 5-$\sigma$
relative to the background. A summary of the data extracted from the
four surveys is given in Table~\ref{tab_surveys}. The
$N_{\rm{CIG}}$ and $N_{\rm{det}}$ entries indicate respectively
the number of galaxies that were covered by each survey and the
number of detections. 
Table~\ref{tab_flux} gives the radio flux density, the radio power, and
a code for the origin of the data according to the following list:
0 for no data, 1 for the original catalog (WENSS, NVSS, GB6), 2 for this work and 3 for NED.
Reference code 4 indicates FIRST detection without a corresponding
NVSS detection. 
 Upper limits are indicated by negative
values. 
 1420 MHz FIRST measures follow in parentheses the NVSS values. 
Fluxes of the 8 galaxies detected with FIRST and
undetected with the NVSS are listed in   Table~\ref{tab_flux} but are
not used in the analysis as explained in Sect.~\ref{secFIRST}. 
Radio power $P$ was computed as
$\log(P) = \log(\frac{S}{\rm{mJy}})+2\log(\frac{D}{\rm{Mpc}})+17.08$.
We used distances $D$ from Verdes-Montenegro et  al. (\cite{verdes05}) with a
Hubble constant of $H_0 = 75$ km\,s$^{-1}$\,Mpc$^{-1}$ updated 
with the newly available data from the bibliography.
The small number of sources without radio power values lack
redshifts. The next
subsections describe  the four surveys used to characterize the
radio continuum properties of our sample.


We derived radio fluxes or upper limits for all galaxies in the CIG 
but we focus our statistical analysis on the complete AMIGA subsample 
described in Verdes-Montenegro et al. (\cite{verdes05}). 
We use in the present work
the same selection as described in Lisenfeld et al. (\cite{lisenfeld07}) (their Sect.~4.1):
(i) The subsample  contains
galaxies with corrected Zwicky magnitudes in the range 11.0--15.0
for which we found $<V/V_m> = 0.43 $, indicating 80--90\% completeness.
(ii)  Morphological revision of the sample, described in Sulentic et al. (\cite{sulentic06}),
identified 32 galaxies that are probably not isolated in the sense that they
might involve isolated interacting pairs and/or multiplets.  These galaxies
are excluded from the most isolated sample studied further here.
(iii) We excluded two nearby dwarf ellipticals (CIG 663 $\equiv$ Ursa Minor and
CIG 802 $\equiv$ Draco) for which we only have upper limits for the radio fluxes,
and the inferred radio luminosities are very low.
This subsample ($N = 719$ galaxies)
will be referred to hereafter as the complete (AMIGA) subsample.

\begin{table}
\caption{Compiled data for the different radio continuum surveys.}
\label{tab_surveys}
\begin{tabular}{llllll}
\hline
\hline
Survey  &  Frequency & Resol. & 5-$\sigma$ &  $N_{\rm{CIG}}^{1}$ & $N_{\rm{det}}^{1}$  \\
    &  (MHz)    & (arcmin) & (mJy)  &  &  \\
\hline
WENSS  &  325 \& 352 &  $\sim 1$ & $\sim 18$  & 405/278 & 49/37\\
NVSS   &  1420  &  $\sim 0.8$  & $\sim 2$ & 1050/719 &  374/311\\
FIRST  &  1420  &  $\sim 0.08$ & $\sim 1$ & 560/360 & 81/58\\
GB6    &  4850  &  $\sim 3.5$   & $\sim 18$ &  1017/691 & 32/12  \\
\hline
\end{tabular}
\begin{list}{}{}
\item[$^{\rm 1}$] Total number of
galaxies with radio data  ($N_{\rm{CIG}}$) and with radio detection
($N_{\rm{det}}$), followed by the number of galaxies belonging to the
complete subsample of $N=719$ galaxies.
\end{list}
\end{table}

\subsection{WENSS}

The WENSS (Rengelink et al. \cite{rengelink97}) survey was carried out with the Westerbork
Synthesis Radio Telescope at 325MHz, except for objects with  $\delta
> 30\degr$ that were observed at 352 MHz. Spatial resolution
is $ \theta_{\rm{WENSS}} \simeq 54\arcsec\times
54\arcsec{\textrm{cosec}}(\delta)$ with limiting flux sensitivity of
about 18 mJy (5-$\sigma$). Astrometric accuracy was 1\farcs5 for
strong sources ($S > 150$ mJy). We adopted a distance tolerance 
of 35\arcsec\ which was chosen taking into account a conservative astrometry accuracy for
weak sources of $\approx \theta_{\rm{WENSS}}/5$ and adopting  a
confidence radius $\sim 3$ times that value. We found 39 WENSS
detections. Five more detections were included from the 352 MHz
polar catalog (CIG 66, 155, 363, 875 and 890).\\

We also downloaded and reprocessed source extractions for other
galaxies from the CIG that are within in the survey area. Flux
calibrations using the integrated SExtractor fluxes of catalog
sources  show no significant differences from the WENSS
catalog values. A total of 49 sources were detected and 404 upper
limits derived. Five weak sources were not listed in the WENSS catalog (CIG
336, 355, 663, 676 and 862)   but are
detected at a 5-$\sigma$ level using SExtractor, showing 
a very weak emission ($<18$ mJy). The total flux  for the extended source CIG 442 
was
underestimated and  recomputed
taking extended emission into account. The astrometry accuracy is
very high, the mean difference with respect to the CIG position
being  $(\alpha,\delta) =  
(0\farcs08,0\farcs34)$. The dispersion in the $\alpha$ and $\delta$
differences are respectively 6\farcs4 and 9\farcs3 which is about
6--9 times smaller than the WENSS spatial resolution. 



\subsection{NVSS}

The NVSS survey  (Condon et al. \cite{condon98}) was made with the VLA at 1.4
GHz. Spatial resolution was $45\arcsec \times 45\arcsec$ for North
of $\delta > -40\degr$, covering the full CIG declination range.
Survey sensitivity is about 2 mJy implying  that nearly all sources
detected with  WENSS (assuming spectral index $\alpha^{1.4}_{0.3} > -1.5$)
will be detected by NVSS. Source positions in the catalog are given
with an accuracy ranging from $<1\arcsec$ for flux density $S>15$
mJy to about $7\arcsec$ for $S<2.5$ mJy. 
We found 359 CIG sources  in the NVSS catalog. Source extraction
using SExtractor with a 5-$\sigma$ threshold yielded 368 detections,
where 9 are not in the  NVSS catalog or were  recalibrated in
cases of extended or weak sources (CIG  324, 436, 442, 543, 559,
837, 869, 950, 988). CIG 199 was excluded from the catalog because
of corrupted data.

\begin{figure}
\resizebox{8.5cm}{8.5cm}{\includegraphics{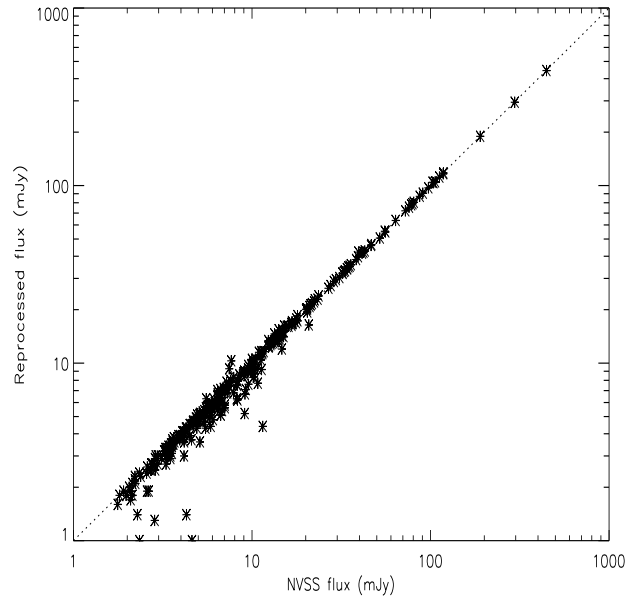}}
\caption{Comparison of the NVSS catalog flux (abscissa) at 1.4 GHz  and the reprocessed fluxes
for the CIG sources (ordinates).}
\label{nvss_calib}
\end{figure}

Comparison of fluxes derived from the NVSS catalog and our
reprocessing 
(see Fig.~\ref{nvss_calib}) 
indicates a high degree of
concordance  except for  an increasing dispersion and
underestimation towards fainter fluxes. These small differences are
probably due to the bias corrections applied to fluxes in the NVSS
catalog and which were not fully taken into account in the
reprocessed fluxes. This difference in calibration affects only the
few sources not listed in the NVSS catalog. In the case of very
extended sources ($> 5\farcm5$; CIG 197, 435, 447, 461, 469, 610) we
have adopted fluxes given in the literature (NED). 
In these cases, the total
flux was computed from a complete integration over the galaxy and
not only from some isolated peaks, as given by the NVSS for the
large and nearby galaxies. A final total of 374 galaxies are
included with 1.4 GHz radio continuum detections. 
The mean difference 
in $\alpha$
and $\delta$ between the NVSS and our catalog positions
are  small $(-0\farcs83,-0\farcs02)$.
The dispersion in the $\alpha$ and $\delta$ differences are
respectively 6\farcs9 and 7\farcs2, which is comparable to the
WENSS astrometry given the similar spatial resolution.

\subsection{FIRST\label{secFIRST}}

FIRST (Becker et al. \cite{becker95}) 
was designed to produce the radio equivalent of the Palomar
Observatory Sky Survey over 10\,000 square degrees of the North
Galactic Cap. The spatial resolution was 5\arcsec. At the 1 mJy
source detection threshold there are $\sim 90$ sources per square degree,
$\sim 35$\% of which show resolved structure on scales from 2\arcsec\ to 30\arcsec.
We cross-correlated the FIRST catalog with our revised optical
positions, yielding 81 source detections. 
The number of detections
was low and the data could not be used for statistical purposes. They 
were however very useful for a  comparison with NVSS data which  allowed us to
infer the relative strengths of disk vs. nuclear emission in many
galaxies. Mean positional differences are small
$(-0\farcs22,-0\farcs07)$ with a dispersion of 1\farcs6 and 2\farcs0,
respectively.
We found 8 galaxies
that were detected in FIRST but not in the NVSS (CIG 236, 238, 258, 364, 544, 618,
678 and 749). Normally we expect the NVSS flux to be stronger than
the FIRST values because FIRST is only sensitive to compact (high
spatial frequency) radio emission. These sources showed fluxes near
the NVSS/FIRST detection limits and were not 
further considered in our analysis.


\subsection{GB6}

The GB6 survey is a combination of two sets of observations made at
the Green Bank telescope in 1986 and 1987 (Gregory et al. \cite{gregory96}). The
sensitivity of the survey is similar to WENSS ($\sim 18$ mJy) but
the spatial resolution is considerably lower (3\farcm5).
Identification of CIG detections in the GB6 catalog required a
threshold distance tolerance of about 1/2 of the FWHM, i.e. 85\arcsec\
which results in 17 detections. SExtractor source extraction with a
detection threshold of 5-$\sigma$ yields  42 detections within the
85\arcsec confidence radius. Comparison of our fluxes 
with GB6 catalog values  indicates a poorer
correlation. Discrepancies  mainly arise from extended sources. The
large spatial resolution compared with the galaxy diameters
motivated a visual check of all detections in order to remove
possible cases of confusion.
A  total of 32 galaxies were retained in our GB6 detection list, with
others determined to be fore-/back-ground.


\begin{table*}
\caption{Radio flux density and power 325 MHz, 1.4 GHz and 4.8 GHz for the CIG
  sample$^{1}$. }
\begin{tabular}{rrrrrrrrrrrrr}
\hline
\hline
  & \multicolumn{4}{c}{ 325 MHz} & \multicolumn{4}{c}{1.4 GHz} & \multicolumn{4}{c}{4.8 GHz} \\
\multicolumn{1}{c}{CIG} &
\multicolumn{1}{c}{Flux}  & 
\multicolumn{1}{c}{Error}&  
\multicolumn{1}{c}{Power} & 
\multicolumn{1}{c}{Ref.}   &
\multicolumn{1}{c}{Flux}   &
\multicolumn{1}{c}{Error}   &
\multicolumn{1}{c}{Power}   &
\multicolumn{1}{c}{Ref.}   &
\multicolumn{1}{c}{Flux}   &
\multicolumn{1}{c}{Error}   &
\multicolumn{1}{c}{Power}   &
\multicolumn{1}{c}{Ref.} \\
 & (mJy) & (mJy)  & (W\,Hz$^{-1}$) &   & (mJy) & (mJy)  & (W\,Hz$^{-1}$) &   & (mJy) & (mJy)  & (W\,Hz$^{-1}$) &   \\
\hline
1  &  0.00  &  0.00 & 0.000  & 0  &   7.70  &   0.72 & 21.894 & 1 & $-$18.00  &  0.00  & $-$22.262  & 1  \\
2  & $-$18.00 &  0.00 & $-$22.228 & 1 &  $-$2.00 &   0.00 & $-$21.274 &  1&  $-$18.00 &  0.00 &  $-$22.228 & 1 \\
3  & $-$18.00 &  0.00 &  0.000 & 1  &  $-$2.00 &   0.00 &  0.000 & 1 &  $-$18.00 &   0.00 &  0.000 & 1 \\
4  &  0.00  & 0.00  & 0.000 & 0   & 31.70  &  1.07 & 21.412 & 1  & $-$18.00  &  0.00 & $-$21.166 & 1 \\
5  &  0.00  & 0.00  & 0.000 & 0   & $-$2.00  &  0.00 & $-$21.380 & 1 &  $-$18.00 &   0.00 & $-$22.335 & 1 \\
6  &  0.00  & 0.00  & 0.000 & 0   &  6.10  &  0.66 & 21.356  &1  & $-$18.00  &  0.00 & $-$21.826 & 1 \\
7  &  0.00  & 0.00  & 0.000 & 0   & $-$2.00  &  0.00 & $-$21.815 & 1 &  $-$18.00 &   0.00 &  $-$22.769 & 1 \\
8  &  0.00  & 0.00  & 0.000 & 0   & $-$2.00  &  0.00 & $-$21.183 & 1 &  $-$18.00 &   0.00 & $-$22.137 & 1 \\
9  & 0.00   & 0.00  & 0.000 & 0   &  2.40  &  0.79 & 21.526 & 1  & $-$18.00  &  0.00 & $-$22.402 & 1 \\
10 & $-$18.00 &  0.00 & $-$21.852 & 1 &  $-$2.00 &    0.00 & $-$20.897 & 1 &
$-$18.00 &   0.00 & $-$21.852 & 1 \\
\ldots&\ldots&\ldots&\ldots&\ldots&\ldots&\ldots&\ldots&\ldots&\ldots&\ldots&\ldots&\ldots\\
\hline
\end{tabular}
\label{tab_flux}
\begin{list}{}{}
\item[$^{\rm 1}$] Full table available in electronic form at the CDS web site
 {\tt http://cdsweb.u-strasbg.fr}
and from {\tt http://www.iaa.es/AMIGA.html}.
\end{list}
\end{table*}


\section{Radio characteristics\label{sectRadioC}}

\subsection{Detection rate}

\begin{figure}
\resizebox{8.5cm}{7.5cm}{\includegraphics{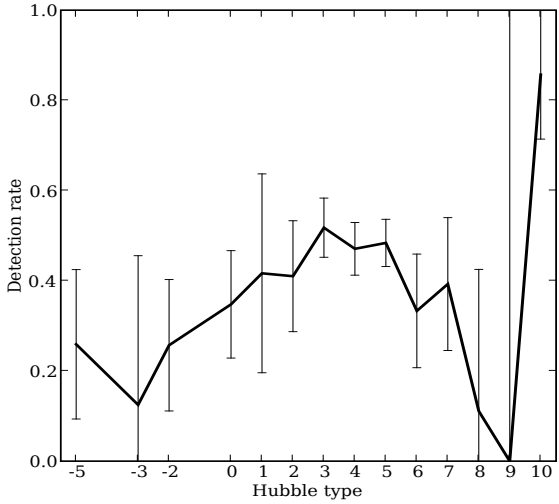}}
\caption{Detection rate at  1.4 GHz with NVSS as a function of Hubble type.}
\label{fig_det-morpho}
\end{figure}

Figure~\ref{fig_det-morpho} shows the NVSS radio detection fraction 
as a function
of morphological type (see Sulentic et al. \cite{sulentic06} for a description of the morphological types). 
It peaks in the Sb--Sc range (type $T=3$--5) which is the core of 
the AMIGA sample, comprising fully two thirds of our sample  (see
Verdes-Montenegro et al. \cite{verdes05}).
Assuming that the radio continuum emission  is driven by star formation
this is not surprising because these galaxies show the highest average
star formation rates (Kennicutt et al. \cite{kennicutt87}). This is partially 
offset by the higher molecular/neutral gas
fraction found in earlier types (Young \& Knezek \cite{young89}) 
which may explain why the
detection fraction drops so slowly towards types earlier than $T=3$ (Sb). 
This can be contrasted to the much more rapid drop in the FIR
detection fraction found in Lisenfeld et al. (\cite{lisenfeld07}). Alternative
explanations for the relatively high detection fraction of earlier types
include: 1) Sb--c spirals misclassied as Sa-ab ($T=1$--2) due to low resolution
imagery; 2) spirals misclassified as E/S0 ($T= -5$ to $-$2); and  3) the growing influence  of
radio emission unrelated to star formation in {\it bona fide} early types.
The large fluctuation among the very late types likely involves small
sample statistics.

\begin{figure}
\resizebox{8.5cm}{8.5cm}{\includegraphics{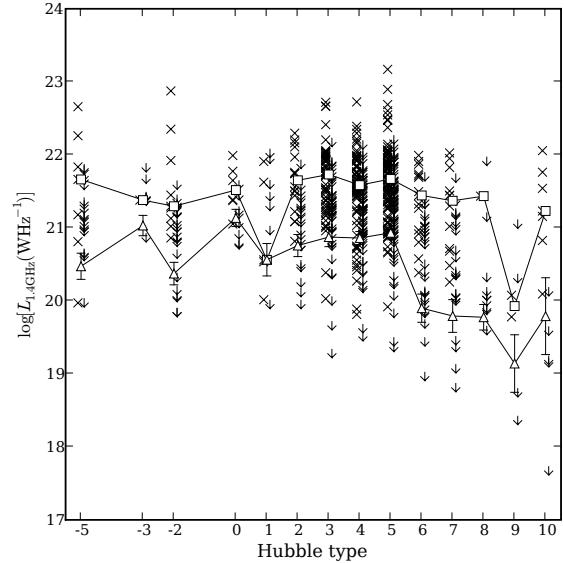}}
\caption{ Distribution of radio luminosity at 1.4 GHz 
as a function of Hubble type. Only
 detections are shown. The open triangles give the mean value of $\log
 (L_{1.4{\rm{GHz}}}$) for
       each Hubble type, calculated with ASURV and taking the upper limits into
 account. The open squares are the median values for the detections
 only.}
\label{fig_lum_t}
\end{figure}

\subsection{Luminosity distribution}

In analogy to earlier papers in this series,  Fig.~\ref{fig_lum_t}
 shows the
distribution of radio luminosities as a function of type for all
detections. In addition, we plot mean values including upper
limits which were derived using the Kaplan-Meier
survival analysis and calculated with the ASURV package (Lavalley et al. \cite{lavalley92}). 
Since the radio detection fraction is less than 50\% for 
all types  the mean values lie quite low relative to the detections.
Note that the mean of $\log(L)$ can be lower than the median of $\log(L)$. 
The weakest known classical radio loud quasars (FRIIs: Sulentic et al. 
\cite{sulentic03}) show radio luminosities near $\log L_{1.4{\rm{GHz}}} = 23.5$ 
and the strongest sources in this sample are 0.5 dex
below, with the majority of detected sources 1.5--2 dex below. With the
exception of a few of the strongest sources all luminosities are
consistent with nonthermal emission related to  star formation.
Figure~\ref{fig_lum_t} can be compared with  the equivalent for FIR emission (Fig. 5
in Lisenfeld et al. \cite{lisenfeld07}).

\begin{figure}
\resizebox{8.5cm}{7.5cm}{\includegraphics{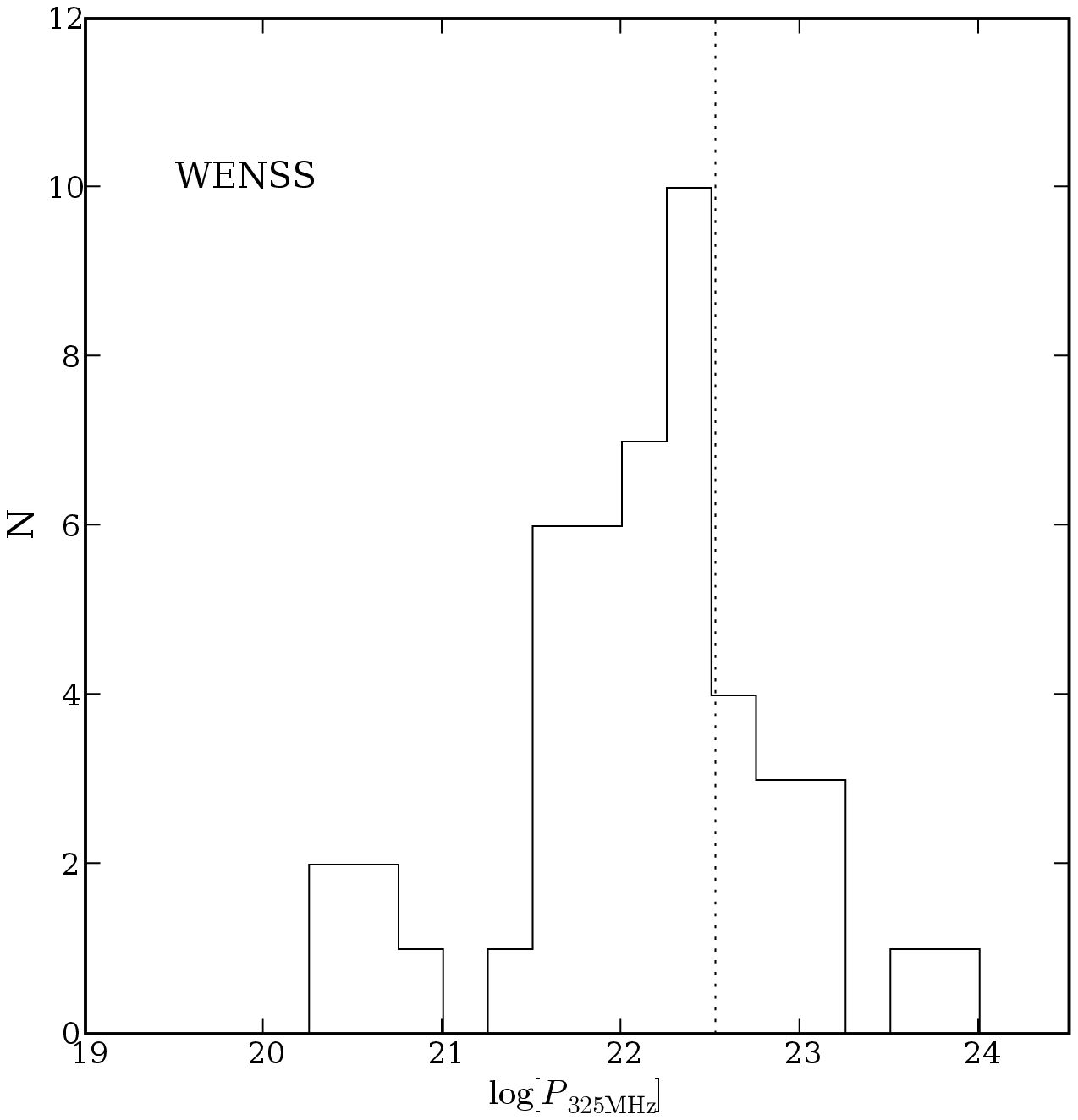}}
\resizebox{8.5cm}{7.5cm}{\includegraphics{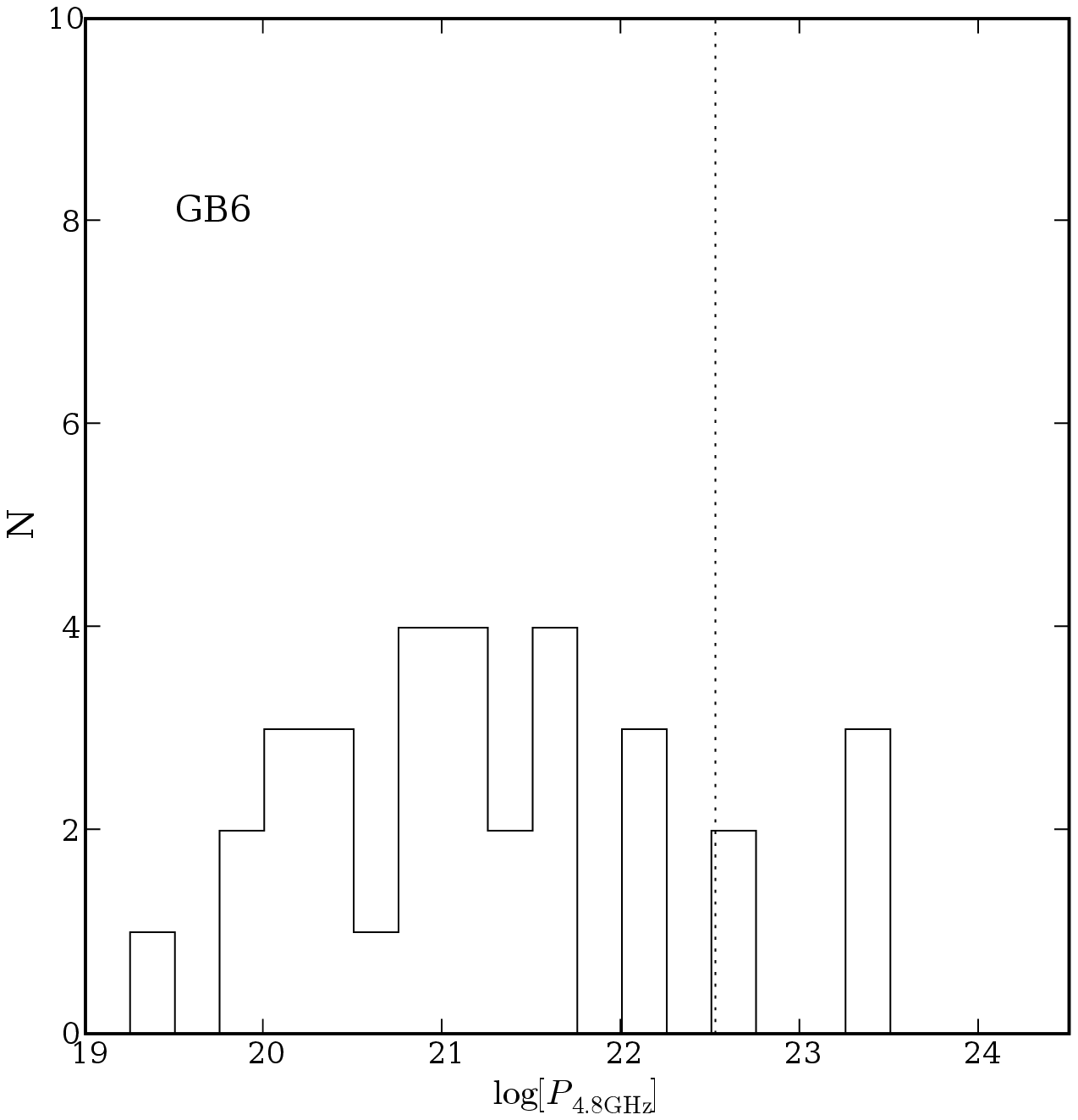}}
\resizebox{8.5cm}{7.5cm}{\includegraphics{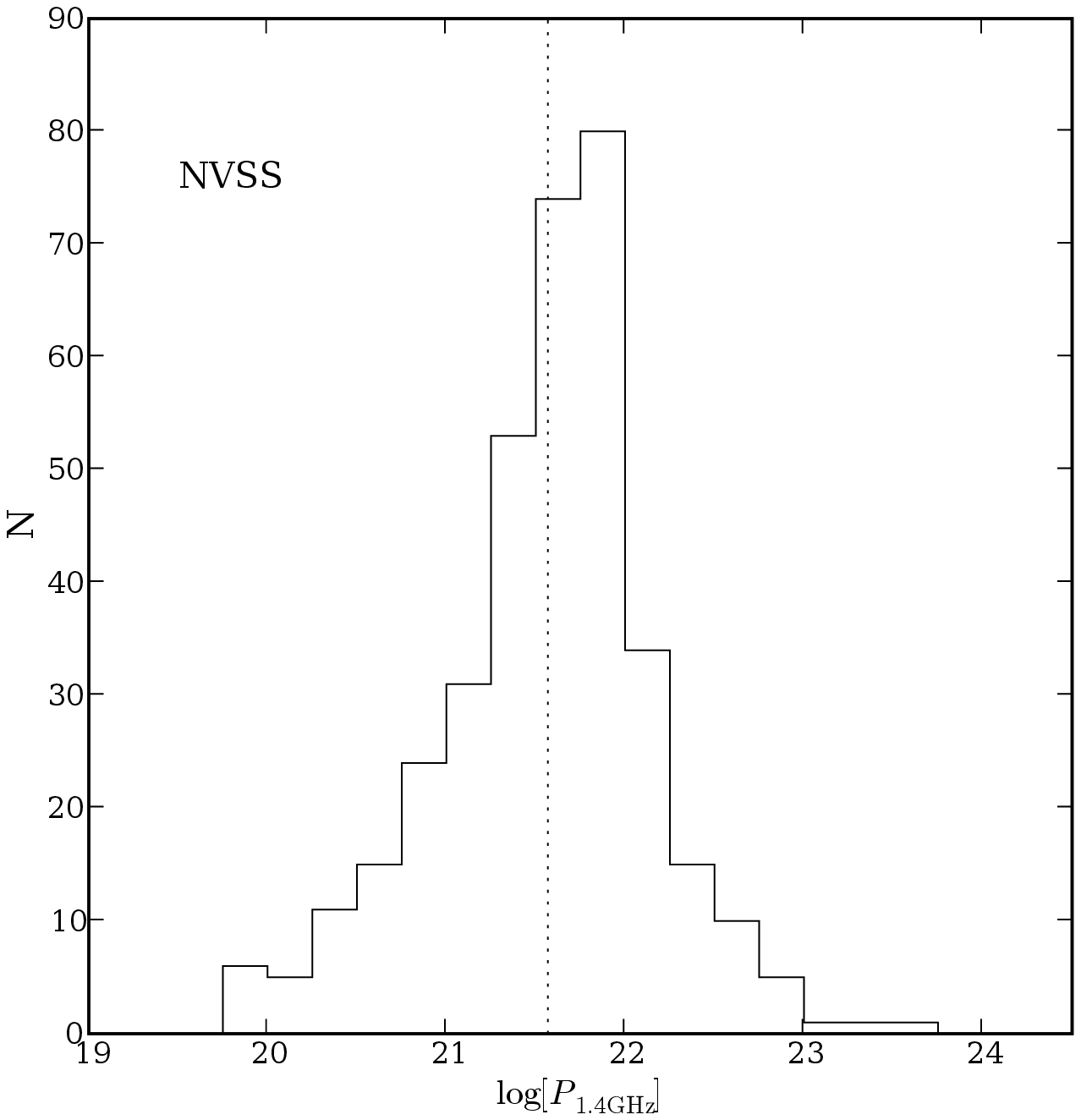}}

\caption{Distribution of the 325 MHz (top), 1.4 GHz (bottom) and 4.8 GHz
  (middle) radio power
for the CIG sample.
The dotted line indicates the completeness for a velocity of 9300 km\,s$^{-1}$ (80\% of the sample).
The $x$-axis indicates the logarithm of the power in W\,Hz$^{-1}$.}
\label{fig_lumfunc}
\end{figure}

The radio power distribution for the three radio bands is shown in Fig.~\ref{fig_lumfunc}.
The vertical dotted line indicates the radio power corresponding to
a completeness higher than 80\% (i.e., more than 80\% of the galaxies are
detected at the corresponding radio frequency). This line is
calculated based on the sensitivity of each survey, combined with
a recession velocity lower than 9300 km\,s$^{-1}$ for 
  80\%  of the CIG galaxies.
We point out that the peak of the distribution for the  NVSS data is above
the level of radio completeness, leading to a better confidence in the radio properties at that frequency.
Table~\ref{tab_radio_lum} shows the mean  value
of the radio power for each frequency using the  Kaplan-Meier survival analysis,
and the median value taking only detections into account. Note that
for the 325 MHz band, the 5 galaxies with data at 352 MHz are included. 
The big difference between the median and mean values 
reflects the large number of upper limits in the samples.
The number of galaxies per frequency band  in the AMIGA sample 
is indicated in the column $N_{\rm{gal}}$,
and the number of detections by $N_{\rm{det}}$.

We compare the radio power distribution of the CIG sample with the one of the star-forming 
galaxies selected from the UGC sample (UGC-SF) in Condon et al. (\cite{condon02}). They studied the radio and star formation 
properties of the entire Uppsala Galaxy Catalog (UGC). They distinguished 
between  star-forming galaxies and  AGN in their study, finding that 
the star-forming galaxy population  may be evolving even at moderately 
low redshift. The AGN have been discarded of the control sample, following their 
classification. We recomputed the radio continuum power and the distances
of the sample by using the same  Hubble constant that for the CIG sample. 
The UGC-SF and the CIG samples are located in a very similar local volume. 

\begin{figure}
\resizebox{8.5cm}{7.5cm}{\includegraphics{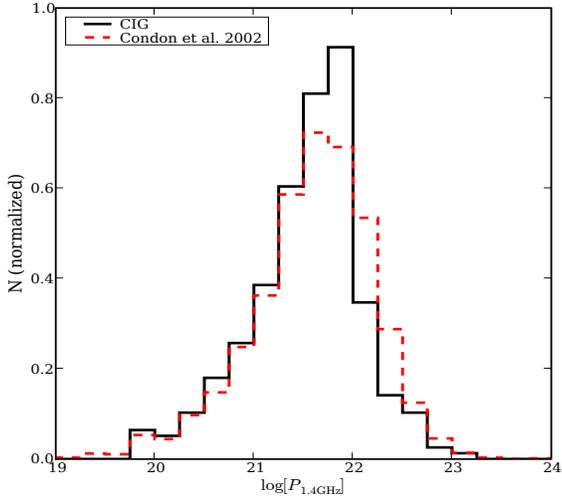}}
\caption{Frequency distribution of the 1.4  GHz radio power 
for the CIG sample (solid line) compared with the frequency distribution of the UGC-SF sample (dashed line) of Condon et al. (\cite{condon02}).}
\label{fig_lumfunc_ugc}
\end{figure}

To compute the mean  value of the 
radio continuum power for the UGC-SF, we used ASURV, as for the CIG sample, by including the detections 
and the upper limits. The corrected $B$ magnitude, retrieved from the LEDA database, was 
used  for the UGC-SF sample.  This  photographic corrected  magnitude is very
similar to the $m_{B{\rm -corr}}$ magnitude of the CIG sample (see Fig.~\ref{fig_lb-cig_condon}).
To be consistent with the CIG sample, we applied
a magnitude cut-off at $m_B =15$ (see Verdes-Montenegro et al. \cite{verdes05}). Using the survival analysis package ASURV, the mean of $\log (P) = 21.61$
 appears significantly higher than for the CIG sample (20.11), as shown in Table~\ref{tab_radio_lum}. The median values are computed  using the detections.  
The difference comes mainly from the upper-limits which are a much 
larger fraction in the CIG sample than in the UGC-SF sample (see Table~\ref{tab_radio_lum}) with a detection rate of 43\% and  90\%  resp. 
for the CIG and UGC-SF samples.
Moreover the Fig.~\ref{fig_lumfunc_ugc}, showing the $\log (P)$ distribution for both samples, indicates
that the UGC-SF sample has a larger population with high radio continuum power ($\log (P) > 22$).

\begin{table}
\title
\caption{Average properties of the radio power, $\log (P)$, for galaxies in the complete AMIGA sample and for the 1.4 GHz radiocontinuum emission of the UGC-SF sample (Condon et al. \cite{condon02}).}
\begin{tabular}{lllll}
\hline
\hline
Frequency  &  Mean $\log(P)$  & Median $\log(P)$ & $N_{\rm{gal}}$ & $N_{\rm{det}}$ \\
       & (W\,Hz$^{-1}$)  & (W\,Hz$^{-1}$) &\\
\hline
CIG/325 MHz    &  20.03$\pm 0.11$ & 22.24 &    278  & 37 \\
CIG/1.4 GHz    &  20.11$\pm 0.10$   & 21.59  &   719   & 311  \\
CIG/4.8 GHz    &  18.54$\pm 0.11$  & 21.53 &   691   & 12\\
UGC-SF/1.4 GHz           & 21.61$\pm 0.62$  & 21.65  & 3136  & 2815 \\
\hline
\end{tabular}
\label{tab_radio_lum}
\end{table}

\subsection{Radio vs. optical properties}

Since our sample involves many of the most isolated galaxies (see Verley et al.
\cite{verley07a}, \cite{verley07b}) in the
local Universe and, if radio emission is enhanced by interactions,
then we expect it to show very  passive radio continuum properties 
relative to almost any other local galaxy sample. 
Figure~\ref{fig_mb_R} compares the radio/optical flux density ratio 
$R$ defined as:

\begin{equation}
R~=~S_{1.4{\rm{GHz}}}\times10^{0.4(m_B\,-14.2)}
\end{equation}

\noindent where $m_B$ is the apparent $B$ magnitude of the galaxy and
$S_{1.4{\rm{GHz}}}$ the flux density (in Jy) at 1.4 GHz. The definition is
similar to the one used by Condon (\cite{condon80}) (the optical magnitude was
normalized to 15.0 in that study). 

\subsubsection{Comparison with KISS}
We chose the value $m_B=14.2$ for
a better comparison with a sample of emission line galaxies (Van Duyne et 
al. \cite{vanduyne04}) which involved  207 emission-line galaxy candidates
from the KPNO International Spectroscopic Survey (KISS) with NVSS or 
FIRST detections. Figure~\ref{fig_histoR} shows the distribution of $R$ values 
for detections in our sample,  with E/S0 and spiral galaxies indicated
separately. $R$ shows a median value of 4.48 which is likely overestimated due
to the
bias introduced by the $\sim 2$ mJy flux limit of the NVSS survey (as
shown in Fig.~\ref{fig_mb_R} by the dashed  line).

As expected, we find the galaxies in our sample to be very
radio quiet relative to their optical luminosities. Generally, high
values of $R$ ($R > 100$, e.g. Van Duyne et al. \cite{vanduyne04}) indicate
the presence of a radio quasar while values between 10 and 100 can arise
from a mix of Seyfert, LINER and starburst activity. Dusty galaxies with 
above average internal extinction can  also boost some normal galaxies 
into this range.  The dominance of low-$R$ galaxies ($R < 10$) in  our 
sample indicates that the radio emission from star formation 
is responsible for the majority of the detections. 
The KISS sample (Van Duyne et al. \cite{vanduyne04}) represents a good
counterpoint with most of the galaxies showing $R > 10$. 
Only three galaxies in our sample show $R >100$ (CIG 187, 349 and
836). CIG  187 has  an underlying background source while CIG 349 is
classified as LINER/Sy1.5 and is almost certainly an interacting/merger system. 
 CIG 836 is a relatively high redshift E galaxy 
($V_{\rm R}$ = 14\,900 km\,s$^{-1}$) and is outside our complete sample. 17\% of the 
Van Duyne et al. (\cite{vanduyne04}) sample show $R>100$.  If we consider the 11 galaxies 
with $R > 30$ in our sample, we find: a) at least one background source; b) 3--5 
interacting sources; and c) 4--5 AGN/LINER galaxies. 
%
%
Isolated examples of 
category c) may represent rare examples of self stimulated AGN activity (via bars?)
or cases where stimulation by an accretion event left no morphological signature.

\subsubsection{Comparison with UGC-SF}

\begin{figure}
\resizebox{8.5cm}{7.5cm}{\includegraphics{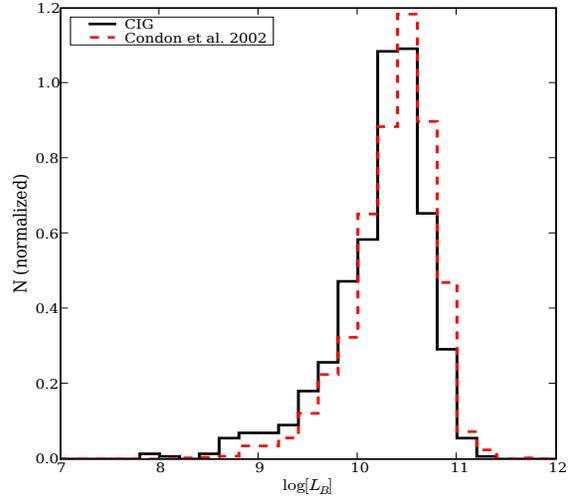}}
\caption{Optical luminosity distribution in $B$ for the CIG (solid line) and the UGC-SF (dashed line) sample.}
\label{fig_lb-cig_condon}
\end{figure}

\begin{figure}
\resizebox{8.5cm}{8.5cm}{\includegraphics{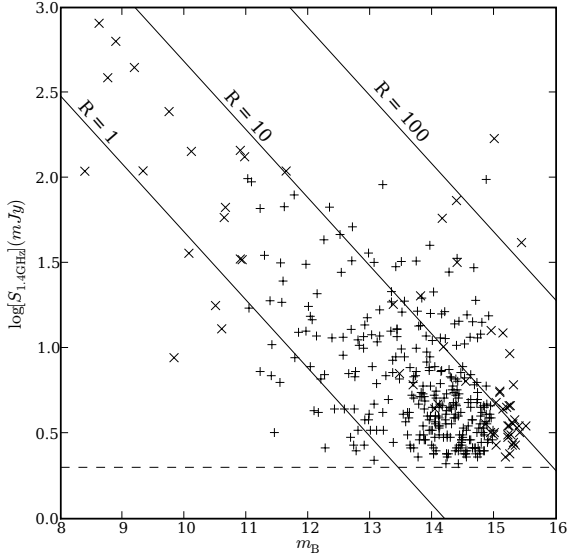}}
\caption{Radio flux density at 1.4 GHz (mJy) vs. apparent
$B$ magnitude. The lines represent constant radio-to-optical ratio  $R$. The dashed line shows
the value of $R$ corresponding to the NVSS sensitivity  level (2 mJy). In order not to overload the figure,
we show only detections. The radio continuum upper limits were however
taken into account in the calculation of the mean values.
The plus signs denote galaxies within the complete AMIGA subsample and
the crosses indicate galaxies outside this subsample.
}
\label{fig_mb_R}
\end{figure}

We compare the $R$ distribution of the CIG with the one of the UGC-SF sample of Condon et al. (\cite{condon02}).
Figure~\ref{fig_histoR} shows little differences between the 
two samples: the $R$ distribution of the CIG is sharper than the one of the UGC-SF sample. This later has 
more important wings, especially an excess for the large $R$ ($\log (R) > 0.8$). The median value of 4.38 
for $R$ of the UGC-SF sample is just slightly larger than the one of the CIG sample (4.19). Figure~\ref{fig_lb-cig_condon}
shows the optical luminosity $L_B$ distribution for the CIG and the UGC-SF sample where it appears that  the UGC-SF sample is slightly brighter in the 
$B$ band than the CIG sample. The similar value of $R$ between both samples is consistent with the higher radiocontinuum and $L_B$ 
for the UGC-SF than for the CIG. The difference for the star formation activity
is minimal between both samples, nevertheless the UGC-SF sample has a larger fraction of galaxies with more 
intense star formation per optical luminosity unit.

\subsubsection{Early types}

\begin{figure}
\resizebox{8.5cm}{7.5cm}{\includegraphics{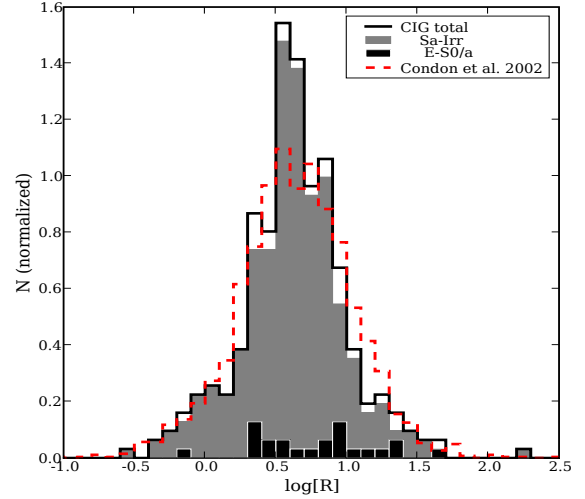}}
\caption{Distribution of the radio-to-optical ratio $R$ (logarithmic scale)  for the CIG sample (solid line). The elliptical and spiral
galaxy distribution are represented respectively by black and grey shaded areas.The dashed line shows the $R$ distribution for the UGC-SF sample (Condon et al. \cite{condon02}). }
\label{fig_histoR}
\end{figure}

Our sample contains a relatively small fraction of early-type E/S0
galaxies ($\sim$~14\%) reflecting the expectation of the morphology-density 
relation. In principle AMIGA provides the last point on the
low density end of that correlation. The low number of early-types
precludes a statistically meaningful comparison of mean $R$ values for
early- and late-type subsamples. Figure~\ref{fig_histoR} suggests a
slight difference between their $R$ distributions, with median values
$R=7.7$ and 4.4 for early- and late-type galaxies, respectively. The $R$
distribution for E/S0 galaxies shows two peaks which may reflect a
true early-type population and a second population of spirals
misclassified as E/S0. If both peaks represent early-types then the
higher $R$ values would be surprising and difficult to understand if
due to star formation activity. We checked whether either the
radio powers or optical magnitudes  for these galaxies had
unusual values and found radio powers in the 10$^{21}$--10$^{23}$ W\,Hz$^{-1}$
range -- overlapping the radio power range of FRI but not 
FRII (broad emission line) quasars. Absolute magnitudes in $B$ 
were generally fainter than $-$21 (except for CIG 893 which was
flagged in our morphological reclassification, in Sulentic et al. \cite{sulentic06},
as a candidate interacting system). If confirmed as E and/or S0 galaxies,  
these very isolated systems would represent one of the most intriguing 
parts of the AMIGA sample.

\begin{figure*}
\resizebox{8.5cm}{8.5cm}{\includegraphics{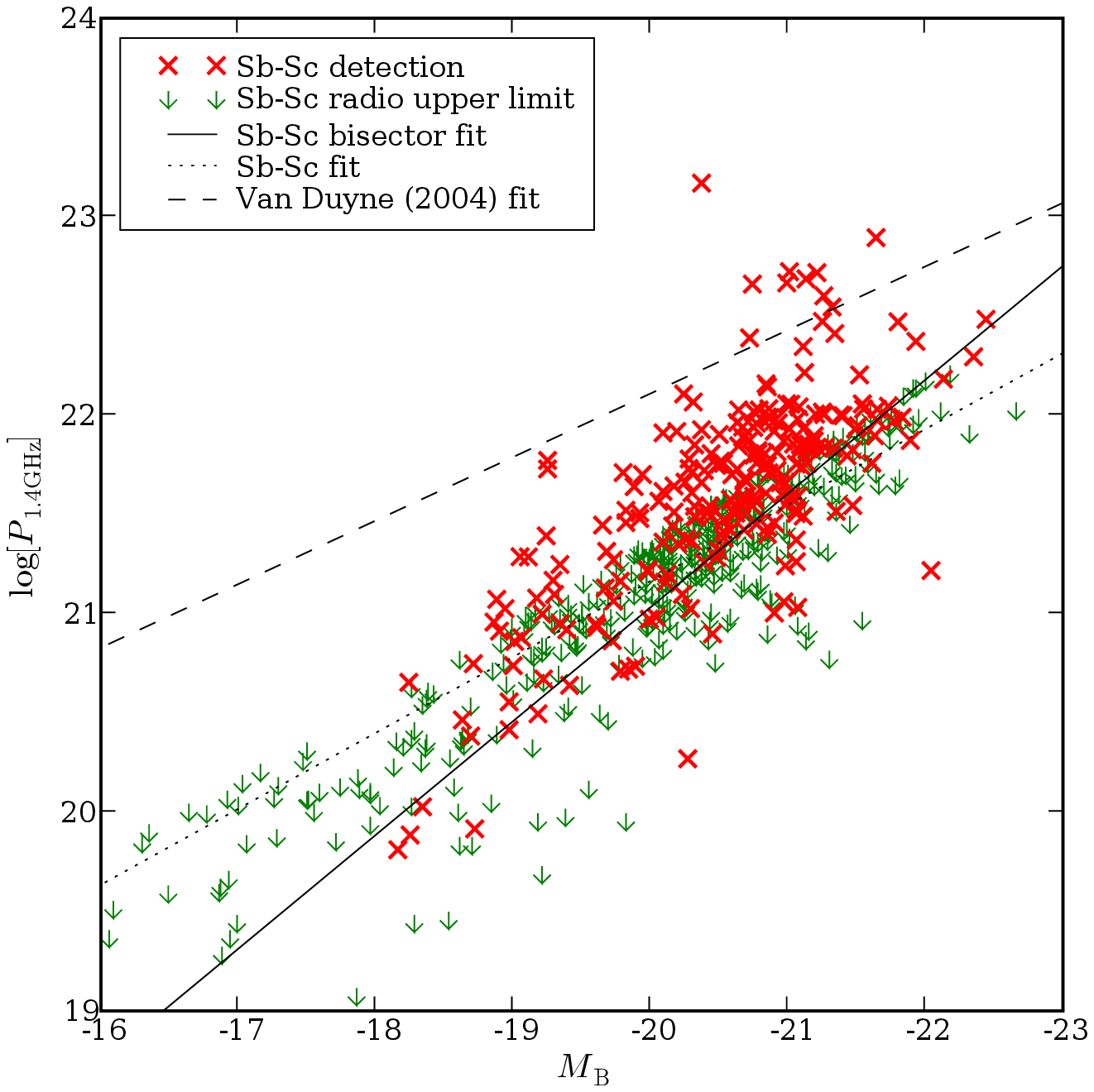}}
\resizebox{8.5cm}{8.5cm}{\includegraphics{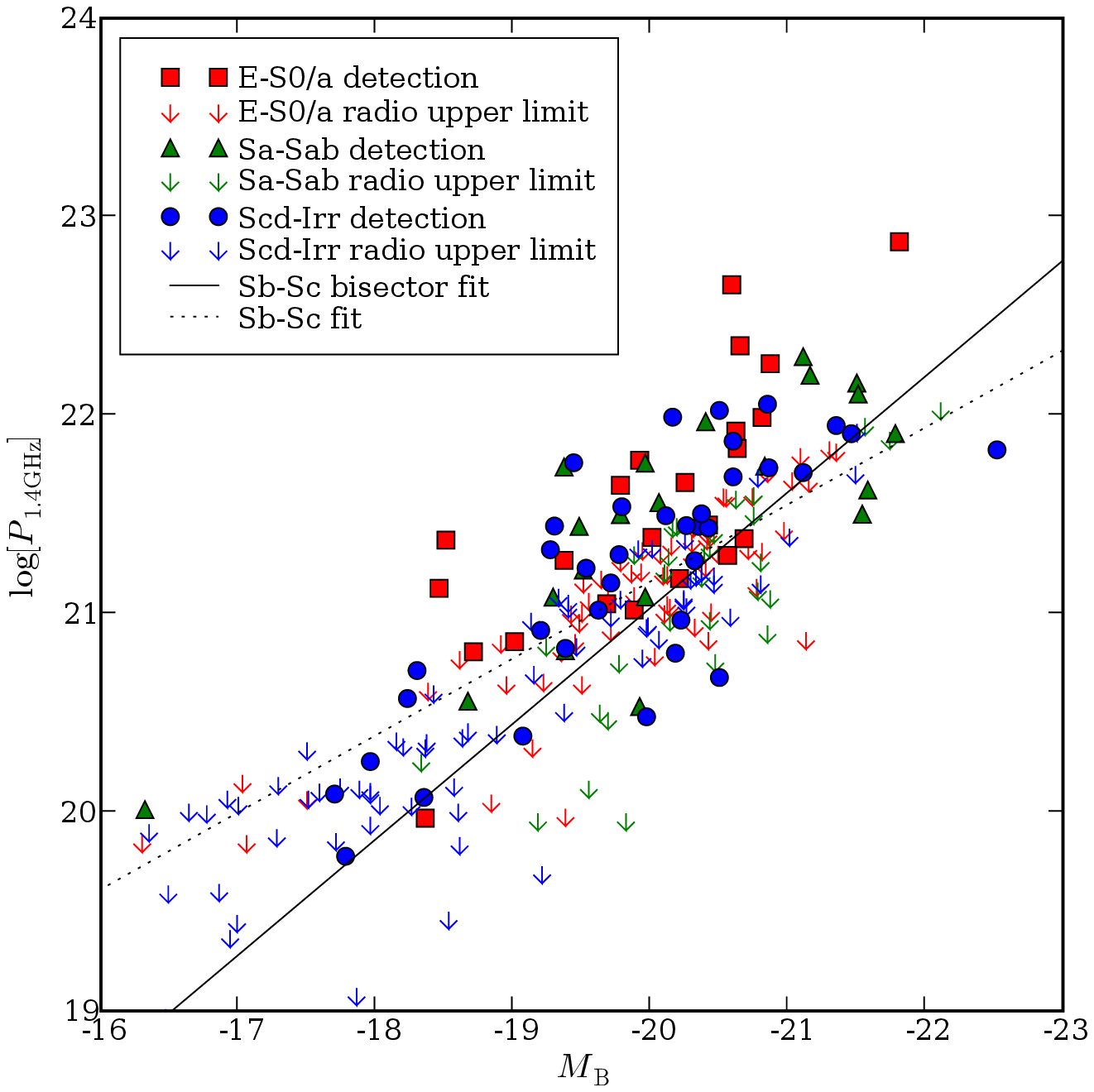}}
\caption{\textbf{Left}: 
Radio power at 1.4 GHz  vs. absolute optical magnitude for the 
Sb--Sc galaxies together with the fitted line obtained using ASURV.
The full line gives the bisector fit. We also plot the fit 
assumming $M_B$ as the independent variable (dotted line) for direct comparison with 
the fit (dashed line) for the KISS starburst galaxies in Van
Duyne et al. (2004).
Detections are indicated by crosses while upper limits in radio
emission by arrows. 
\textbf{Right}: Radio power at 1.4 GHz  vs. absolute optical magnitude for the 
detected E-S0 (filled squares), Sa-Sab (filled triangles) and 
Scd-Irr (filled circles) galaxies. The upper limits are indicated with arrows of the same 
color.
The lines correspond to the same fits as in the left panel.
}
\label{fig_mabs_radiopower}
\end{figure*}

\subsection{ Radio-optical luminosity correlations}

The relationship between optical and radio continuum properties of
galaxies in our sample can also be evaluated by studying the
absolute $B$ magnitude vs. log radio power (W\,Hz$^{-1}$) correlation.
The most useful information comes from 
using the 1.4~GHz data where the detection fraction was highest
 (Fig.~\ref{fig_mabs_radiopower}).
We use both detections and upper limits (employing survival analysis
techniques) for calculation of radio continuum properties. The
relation between radio continuum power and absolute $B$ magnitude was
derived using the Schmidt method (see Feigelson \& Nelson \cite{feigelson85}) as
implemented in the ASURV package. A random
distribution of upper limits is required for application of survival
analysis techniques. In our case, the upper limit distribution of the
radio power is assumed random because it was computed from a flux
limited distribution combined with a random distribution of galaxy
distances. We get the following relations for the complete
subsample taking $M_B$ as the independent variable:

\begin{eqnarray}
%
\log(P_{325{\rm{MHz}}}) &  =  & (-0.49 \pm 0.03)M_B+(11.7 \pm 0.6) \\
\log(P_{1.4{\rm{GHz}}}) &  = & (-0.43 \pm 0.02)M_B+(12.4 \pm 0.5) \\
\log(P_{4.8{\rm{GHz}}}) & = & (-0.53 \pm 0.02)M_B+(10.6 \pm 0.5)
\end{eqnarray}
A bisector fit gives:
\begin{eqnarray}
\log(P_{325{\rm{MHz}}}) &  =  & (-0.60 \pm 0.03)M_B+(9.4 \pm 0.5) \\
\log(P_{1.4{\rm{GHz}}}) &  = & (-0.61 \pm 0.02)M_B+(8.7 \pm 0.5) \\
\log(P_{4.8{\rm{GHz}}}) & = & (-0.60 \pm 0.03)M_B+(9.1 \pm 0.6)
\end{eqnarray}

The bisector slope of the radio power versus  optical magnitude correlation
is steeper at 0.3 and 4.8 GHz than at 1.4 GHz, but 
within the errors.
In order to avoid mixing different morphological types, 
and hence different factors contributing to the radio
emission, we derived a fit for the Sb--Sc majority population of 
the AMIGA sample (Fig.~\ref{fig_mabs_radiopower}, left):

\begin{eqnarray}
\log(P_{1.4{\rm{GHz}}}) &  = & (-0.58 \pm 0.02)M_B+(9.4 \pm 0.4)\\
\end{eqnarray}
and taking $M_B$ as independent variable:

\begin{eqnarray}
\log(P_{1.4{\rm{GHz}}}) &  = & (-0.39 \pm 0.02)M_B+(13.4 \pm 0.5)\\
\end{eqnarray}

Figure~\ref{fig_mabs_radiopower} (right) shows earlier and later type galaxies in our 
sample along with best fit linear regression and bisector fits 
to the Sb--c majority. The scatter in these relations (especially the Sb-Sc one)
is still very large and will be explored in more detail in a later paper. 
Examination of the highest points reveals a mix of interacting systems and 
AGN (see Sabater et al. \cite{sabater08}) along with a few expected background sources.
The lowest points tend to belong to highly inclined spirals, likely telling 
us that further refinement of inclination corrections and resultant optical 
luminosity corrections are needed.

It is interesting that almost all radio continuum detections of early-types
lie above the best fit relation (Fig.~\ref{fig_mabs_radiopower}, right). If the spiral majority shows 
radio emission related to star formation, then this offset is likely telling us that
radio emission in early-types has a different origin.
Figure~\ref{fig_mabs_radiopower} (left) shows that the CIG sample has a lower radio
power than Van Duyne et al. (\cite{vanduyne04}) over a similar absolute magnitude
range. This is another manifestation of the strong difference
between an emission line selected sample and our sample of isolated galaxies.
Since no selection was made in the Van Duyne et al. sample in terms of
the environment, it is a reasonable assumption  that it involves both 
interacting systems and galaxies in generally higher density environments. 
The two samples show no sources in common.

\subsection{Comparison between NVSS and  FIRST fluxes}

The NVSS and FIRST surveys were obtained with different VLA
configurations, resulting in different effective resolutions and,
more importantly, in diminished spatial frequency sensitivity
(scales larger than $\sim$ 30\arcsec)  for FIRST.
Comparing NVSS and FIRST observations of the same source then
provides insights about the spatial distribution of emission in a
galaxy. FIRST will largely sample only compact (likely nuclear)
emission while NVSS will detect both nuclear and disk emission from
galaxies in our sample. If low level star formation dominates the
radio emission from most of our galaxies, then we expect NVSS fluxes
to be larger than corresponding FIRST fluxes for our sample. A
strong FIRST detection might indicate an active nucleus; these
sources, expected to be rare in an isolated sample like AMIGA, will
be discussed in a companion paper (Sabater et al. \cite{sabater08}).

\begin{figure}
\resizebox{8.5cm}{7.5cm}{\includegraphics{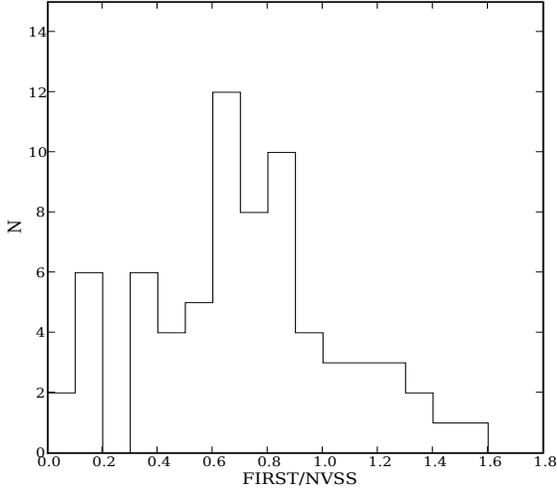}}
\caption{Histogram of the flux ratio between the  FIRST and NVSS detections for the CIGs.}
\label{fig_hist_firstnvss}
\end{figure}

The much smaller FIRST  detection fraction (14\% vs. 35\%
for NVSS) immediately tells us what to expect -- a
sample dominated by disk star formation. FIRST fluxes are smaller
than NVSS values in almost all sources. Most galaxies with a FIRST
detection are Sb--Sc spirals, which represent 2/3 of the entire
isolated population. This detection fraction suggests that more than half of
our NVSS detected galaxies have disk-dominated radio continuum
emission at 1.4~GHz because they are fully attenuated by FIRST.
Figure~\ref{fig_hist_firstnvss} shows the FIRST/NVSS flux ratio
distribution for our sample. 
About 40\% of the galaxies detected by FIRST show core-dominated
emission at 1.4 GHz, which implies $\sim$ 20\% of  core-dominated
galaxies in our sample. The radio continuum emission in
core-dominated galaxies may originate either from nuclear SF or from
an active nucleus. That question is addressed in Sabater et
al. (\cite{sabater08}).

\begin{figure}
\resizebox{8.5cm}{7.5cm}{\includegraphics{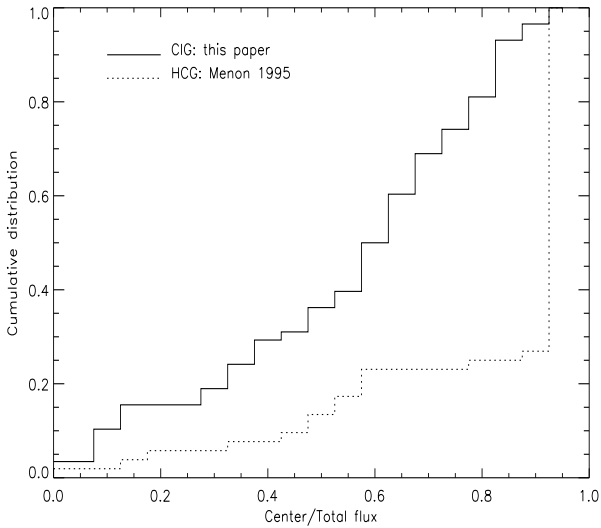}}
\caption{Cumulative distribution of the FIRST/NVSS flux ratio for the CIG (solid line) and the center/total flux ratio (dotted line) from Menon (\cite{menon95}).}
\label{fig_CIG_HCG}
\end{figure}

We note that   Menon (\cite{menon95}), using VLA B- and C-
configurations at 1.4 GHz, found that 46/56 radio detected members
of compact galaxy groups showed compact core radio emission. Figure~\ref{fig_CIG_HCG} compares the cumulative distribution function of
FIRST/NVSS flux ratios for AMIGA and the Menon (\cite{menon95})  compact group
sample -- a striking comparison of nature vs. nurture in the radio
domain. The CDF comparison shows that only 13\% of compact groups
have a ratio lower than 0.5, while 36\% are found for the CIGs.
This shows that isolated galaxies have a more disk-dominated
radio emission than galaxies in groups where extreme interaction
effects are expected.

\section{Radio Luminosity Function (RLF)\label{sectLF}}

In the past few decades many researchers have studied the radio
continuum properties of different samples, including derivations of 
the RLF. A large fraction of these studies have involved flux limited
samples useful for studying the evolution of the RLF.
The 1.4 GHz NRAO VLA Sky Survey (NVSS) showed that radio  evolution is 
largely luminosity independent with strong cosmological evolution at all 
luminosities (Condon \cite{condon84}, \cite{condon89}). Benn et al. (\cite{benn93}) found that 
faint radio sources ($S_{1.4{\rm{GHz}}}< 1$ mJy) are dominated by star-forming 
galaxies with only a small fraction showing Seyfert type emission. 
AGN dominate the bright end of the radio distribution. Galaxies from an 
IRAS flux-limited sample ($S_{60\mu{\rm m}} > 2$ Jy) were cross-matched 
with the NVSS (Yun et al. \cite{yun01}), revealing that FIR-selected galaxies 
can account for the entire population of late-type field galaxies in 
the local Universe. Radio emission in these galaxies is mainly produced 
by star formation. 
One of the most
recent radio continuum studies for a large sample involves the 2
degree Field Galaxy Survey (2dFGSDR2) cross-correlated with the NVSS
for galaxies brighter than $K = 12.75$ (Sadler et al. \cite{sadler02}). Using
2dF spectra, they found that 60\% of the sources involve dominance of 
star formation and 40\% AGN. Interpretation of all of these surveys is 
clouded by the uncertain role of environment. The AMIGA sample has the 
potential to clarify this role by providing a nurture-free 
local sample, large enough to permit evaluation as a function of source morphology
and luminosity.

\subsection{1.4 GHz radio luminosity function}

The local  RLF  of the CIG sample is useful 
in order
to get more insight into the cosmological evolution
of the radio emission in isolated galaxies, as well
as to compare them  with other galaxy populations in the local Universe.
We only compute the local RLF at 1.4 GHz, for which we can get an accurate statistics thanks to
the large number of detections. To compute the RLF we follow the prescription given by Xu \& Sulentic (\cite{xu91}).
Since the CIG sample is optically selected, the RLF is derived from the OLF
and the fractional bivariate function between the radio and optical luminosity. The OLF $\phi(M)$ was estimated by Verdes-Montenegro
et al. (\cite{verdes05}) using the $<V/V_m>$ test (Schmidt \cite{schmidt68}) which indicates a high level of completeness for the CIG.   The RLF is then
calculated for that complete
subsample.

\begin{figure}
\resizebox{8.5cm}{8.5cm}{\includegraphics{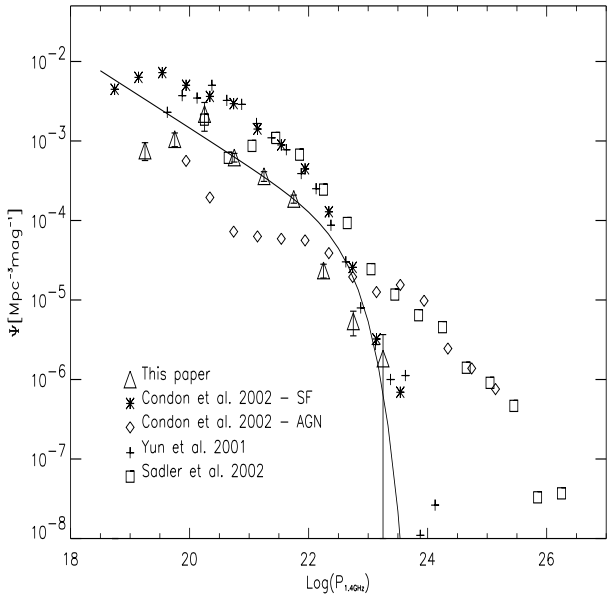}}
\caption{RLF at 1.4 GHz for the CIG sample (triangles) with  1-$\sigma$ error bars. The RLF is shown
in Mpc$^{-3}$ mag$^{-1}$. The
line represents the Schechter fit to the CIG sample,
taking into account only data points with 
$\log(P) \ge  20.75$.
}
\label{fig_RLF}
\end{figure}

The differential RLF $\psi$ gives the number of galaxies per unit volume and per unit $\log P_{1.4{\rm{GHz}}}$
interval. It is derived as follows:
\begin{equation}
\psi (P) ~=~2.5 \Delta M \sum_{i} \Theta(P | M_i) \phi(M_i)
\end{equation}

\noindent where $P = \log (P_{1.4{\rm{GHz}}})$.  $\Theta(P | M_i)$ is the bivariate (optical, radio) luminosity function (BLF).
The factor 2.5 arises since dex(0.4) = 1 magnitude. The BLF is defined as:

\begin{equation}
\Theta(P | M_i) ~=~ \frac{\mathcal{P}(P|M_i)}{\Delta P}
\end{equation}

\noindent where $\Delta P ~=~0.5$ and $\mathcal{P}(P|M_i)$ is the conditional probability for a source with an absolute
magnitude $M$ ($M_i+0.5 \Delta M \geq M > M_i-0.5 \Delta M $) to have the logarithm of its radio
luminosity, $\log (P_{1.4{\rm{GHz}}})$, within the interval [$P-0.5\Delta P,P+0.5\Delta P$].
In order to compute the BLF, we use the Kaplan-Meier estimator to take into account the upper limits of the
radio observations, using the ASURV package. The errors of
$\psi (P)$ are the quadratic sums of the uncertainties of the OLF and the Kaplan-Meier estimator.
The RLF is shown in Fig.~\ref{fig_RLF} and the BLF is listed in Table~\ref{tab_blf}. We point out that the inclusion of the
upper limits in the computation may lead to a different final RLF compared with  other studies, biased due to the use of only
radio continuum detections. The survival analysis was designed to use the maximum of information available from the data and to
provide a closer representation of the ``true'' distribution.
The BLF is dominated by the small number statistics mainly at low optical and radio luminosity, due to the lack
of faint galaxies in the CIG catalog. A clear correlation between the optical and radio luminosity shows up in the
BLF, with the noticeable lack of galaxies with a strong radio continuum emission ($\log(P_{1.4{\rm{GHz}}}) \ge 23.50$).
The RLF at 1.4 GHz is given in Table~\ref{tab_rlf}.

\begin{table*}
\caption{Bivariate (optical,radio) luminosity function for the CIG sample.}
\begin{tabular}{lr|llllllllllllll}
\hline
\hline
 & $\!\!\!\!\!\!M$ & $-$16.25 & $-$16.75 & $-$17.25 & $-$17.75 & $-$18.25 & $-$18.75 & $-$19.25 & $-$19.75 & $-$20.25 & $-$20.75 & $-$21.25 & $-$21.75 & $-$22.25 \\
$\log(P)$ &&&&&&&&&&&&&\\
\hline
18.75 && 0.00   & 0.00   &  0.00   &  0.00   &  0.00   &  0.00   &   0.00  &  0.00  &  0.00  & 0.00  & 0.00  & 0.00  & 0.00 \\
19.25 && 0.00   & 0.00   &  0.00   &  0.81   &  0.31   &   0.83  &  0.00  &  0.00  & 0.00  & 0.00  & 0.00  & 0.00  & 0.00\\
19.75 && 0.00   & 0.00   &  0.00   &  0.81   &  0.94   &  0.28   &   0.63  &  0.40  &  0.00  & 0.00  & 0.00  & 0.00  & 0.00 \\
20.25 && 2.00   & 0.00   &  0.00   &  0.37   &  0.42   &  0.26   &   0.16  &  0.13  &  0.35  & 0.00  & 0.00  & 0.00  & 0.00 \\
20.75 && 0.00   & 0.00   &  0.00   &  0.00   &  0.25   &  0.46   &   0.70  &  0.65  &  0.77  & 0.47  & 0.45  & 0.31  & 0.00 \\
21.25 && 0.00   & 0.00   &  0.00   &  0.00   &  0.08   &  0.17   &   0.39  &  0.57  &  0.47  & 0.61  & 0.34  & 0.00  & 1.07 \\
21.75 && 0.00   & 0.00   &  0.00   &  0.00   &  0.00   &  0.00   &   0.12  &  0.24  &  0.35  & 0.73  & 0.83  & 1.22  & 0.00 \\
22.25 && 0.00   & 0.00   &  0.00   &  0.00   &  0.00   &  0.00   &   0.00  &  0.00  &  0.04  & 0.14  & 0.29  & 0.38  & 0.93 \\
22.75 && 0.00   & 0.00   &  0.00   &  0.00   &  0.00   &  0.00   &   0.00  &  0.00  &  0.00  & 0.04  & 0.10  & 0.09  & 0.00 \\
23.25 && 0.00   & 0.00   &  0.00   &  0.00   &  0.00   &  0.00   &   0.00  &  0.00  &  0.01  & 0.00  & 0.00  & 0.00  & 0.00 \\
\hline
\end{tabular}
\label{tab_blf}
\end{table*}

Figure~\ref{fig_RLF} shows a comparison of the CIG RLF with other samples. 
Except for the AGN sample of Condon et al. (\cite{condon02}), all
samples have a relatively similar RLF at low 
radio power ($P_{1.4{\rm{GHz}}}<10^{22} \mbox{ W\,Hz}^{-1}$). The radio emission of these galaxies
showing a similar contribution to the local density of the RLF is dominated by SF.
At larger radio power, the  2dFGRS sample from Sadler et al. (\cite{sadler02}) and
AGN/UGC  dominate the radio contribution, that is, the radio power is coming mainly from the AGN-powered galaxies.
The CIG RLF decreases below the AGN dominated samples for powers higher than
 $P_{1.4{\rm{GHz}}}\sim10^{23}$ W\,Hz$^{-1}$, in very good coincidence with samples powered by SF. Thus, the  radio
emission from the CIG sample is mainly powered by SF and the AGN contribution is very low, as confirmed
by Sabater et al. (\cite{sabater08}).

The Schechter formalism to describe the luminosity function (Press \& Schechter \cite{press74}) was initially
developed to explain the luminosity function of galaxies based on the cosmological evolution of the
clustering of matter in the Universe. Its application to radio frequencies  (e.g. Yun et al. \cite{yun01};
Serjeant et al. \cite{serjeant02}) is a powerful tool to compare the distribution of various
kinds of radio sources and to parametrize their  evolution. The Schechter function for a 
RLF $\rho (L)$ is the following:

\begin{equation}
\rho(L){\rm d}L~=~\rho_\star \left ( \frac{L}{L_\star} \right )^\alpha \exp \left (-\frac{L}{L_\star} \right )
{\rm d}\left (\frac{L}{L_\star} \right )
\end{equation}

\noindent where $\rho_\star$ and $L_\star$ are the characteristic density and luminosity of the population and $\alpha$
describes the faint-end power-law slope for $L \ll L_\star$. 
Our RLF cannot be well described by a single Schechter function. We therefore
derived a fit only for the high luminosity data points, those with
$\log(P_{1.4{\rm{GHz}}}) \ge 20.75$.
The fitted parameters are given in Table~\ref{tab_schechter_rlf}.  
The fits for Condon et al. (\cite{condon02})  and Yun et al. (\cite{yun01}) have been recomputed for consistency.
The $\chi^2$ of the fit is indicated together with the degree of freedom of the fit (DoF).\\
The $L_\star$ parameter is quite uncertain because of the relative larger errors of the bright end data.
Yun et al. (\cite{yun01}) found that fitting their RLF with a sum of two Schechter functions gives a better fit.
We tried such decomposition for the CIG sample, and the $\chi^2$ was not improved but increased. It shows the limit
of the application of the Schechter formalism  to the RLF,  developed to model the mass distribution
of the sub-structures in the formation of the galaxies.
Yun et al. (\cite{yun01}) noted that their decomposition fits two different galaxy populations: one normal, late-type field
galaxies, and a second composed of starburst and luminous infrared galaxies. It appears that the ``field'' galaxies of the
IRAS-2 Jy sample have a closer  Schechter decomposition to the CIG population than the starburst galaxies with
$L_\star \sim 2\times10^{22}$ W\,Hz$^{-1}$ and  $\alpha\sim -0.63$. The lower power-law slope of the CIG compared to the 
UGC-SF sample indicates that
the CIG sample has a larger population of low  luminosity  galaxies. The same conclusion arises from the lower $L_\star$
value than the IRAS-2 Jy, since they have a comparable power-law slope.


\begin{table}
\caption{RLF at 1.4 GHz for the CIG sample.}
\begin{tabular}{lll}
\hline
\hline
\multicolumn{1}{c}{$\log{P_{1.4{\rm{GHz}}}}$}&
\multicolumn{1}{c}{$\psi$} &
$N$ \\
(W\,Hz$^{-1}$)  &  (Mpc$^{-3}$\,mag$^{-1}$) & \\
\hline
       19.25  &  $1.9(\pm 0.5)\times 10^{-3}$  & 24.6\\
       19.75  &  $2.6(\pm 5.5)\times 10^{-3}$  & 55.9\\
       20.25  &  $5.5(\pm 2.1)\times 10^{-3}$ &52.4\\
       20.75  &  $1.6(\pm 0.2)\times 10^{-3}$  & 188.0 \\
       21.25  &  $9.0(\pm 0.1)\times 10^{-4}$  & 144.6 \\
       21.75  &  $4.7(\pm 0.5)\times 10^{-4}$  &165.7 \\
       22.25  &  $5.9(\pm 0.1)\times 10^{-5}$  &39.8 \\
       22.75  &  $1.3(\pm 0.5)\times 10^{-5}$  & 10.0 \\
       23.25  &  $4.6(\pm 4.6)\times 10^{-6}$  &1.0\\
\hline
\end{tabular}
\label{tab_rlf}
\end{table}

%


\begin{table}
\caption{Schechter function for the RLF at 1.4 GHz. }
\label{tab_schechter_rlf}
\begin{tabular}{lllll}
\hline
\hline
Survey & \multicolumn{1}{c}{$\rho_\star$}   & \multicolumn{1}{c}{$L_\star$}  &
\multicolumn{1}{c}{$\alpha$}   & \multicolumn{1}{c}{$\chi^2$/DoF} \\
       & \multicolumn{1}{c}{($\times 10^{-3}$)}    &
       \multicolumn{1}{c}{($\times 10^{23}$)} & & \\
\hline
CIG  & $34.2(\pm 9.0)$  &  $0.43(\pm 0.23)$  &  $-0.47(\pm 0.06)$ & 0.04/3 \\
UGC-SF$^1$ & $46.8(\pm 31.0)$  &  $0.94(\pm 0.24)$   &  $-0.48(\pm 0.08)$ & 0.1/10 \\
IRAS-2Jy$^2$  &  $19.6(\pm 13.8)$   & $2.30(\pm 1.30)$ & $-0.73(\pm 0.11)$ &
0.8/14 \\
\hline
\end{tabular}
\begin{list}{}{}
\item $^1$ Condon et al. (\cite{condon02}), SF sample.
\item $^2$ Yun et al. (\cite{yun01}).
\end{list}
\end{table}


\subsection{Radio power density}


We chose to compute the total radio power density from the RLF  at 1.4 GHz rather than from the Schechter fit,
since
the parameters of the Schechter fit would give the total power density as
$U_{1.4{\rm{GHz}}} = \rho_\star L_\star \Gamma (2+\alpha) $
for a larger range than the available radio data.
Given the uncertainties on the parameters and the extrapolation of the formula, we preferred to compute $U_{1.4{\rm{GHz}}}$ only on the radio power
range of the RLF.
 The results  for the CIG and the other samples considered from the bibliography are given in
Table~\ref{tab_powerdensity}: the radio power density for each  sample is
computed for a Hubble constant of $H_0 = 75$ km\,s$^{-1}$.
The CIG sample has, by far, the lowest contribution ($4.4\times 10^{18}$ W\,Hz$^{-1}$Mpc$^{-3}$) in the Local Universe
to the radio continuum emission at 1.4 GHz. Its contribution is 30\% of the
total power density  of the IRAS-2 Jy sample
and 6\% compared to the 2dFGRS, which covers a redshift range of $z = 0.005$ to 0.438 and has 60\% of active galaxies and 40\% of star-forming
galaxies (Sadler et al. \cite{sadler02}). The difference of the
total radio luminosity between the UGC and IRAS-2 Jy samples is mainly due to the 
lower faint end of the later one.
It appears clearly that the galaxies hosting an AGN are dominating the sources 
emitting at 1.4 GHz. 
The low contribution of the CIG galaxies to the total radio power density is the combination of
two effects: (i) a slightly lower radio emission of the individual galaxies, in comparison to other environments,
as shown for the mean $\log(L)$  in comparison to the UGC-SF sample of Condon et al. (\cite{condon02}), and  (ii)  a low number density
of isolated galaxies in the local Universe.
\\

\begin{table}
\caption{Radio power density.}
\begin{tabular}{lll}
\hline
\hline
Survey & $U_{1.4{\rm{GHz}}}$  & References \\
 & {\scriptsize 10$^{18}$ W\,Hz$^{-1}$Mpc$^{-3}$ }  & \\
\hline
CIG  & 4.44 $\pm 0.52$  & This paper \\
IRAS-2Jy & 15.60 $\pm 0.25$  & Yun et al. (\cite{yun01}) \\
UGC/NVSS-SF & 17.06 $\pm 0.35$  &  Condon et al. (\cite{condon02})  \\
UGC/NVSS-AGN & 41.88 $\pm 3.47$  & Condon et al. (\cite{condon02})  \\
2dFGRS & 74.76 $\pm 3.59$ & Sadler et al. (\cite{sadler02}) \\
\hline
\end{tabular}
\label{tab_powerdensity}
\end{table}

\section{Conclusions\label{sectDisc}}



This paper is part of a series designed to characterize the
ISM in a sample of the most isolated galaxies in the local Universe.
We are establishing multiwavelength baselines or zero-point levels
for the stellar content and the different components of the ISM
against which samples in different environments can be compared and
interpreted. 
Radio continuum emission that
arises from either star formation or active nuclei should be at the
lowest ``natural'' level in the AMIGA sample. We analyzed radio
continuum emission at three frequencies: 325 MHz, 1.4 and 4.8 GHz using
data from the WENSS, NVSS, FIRST  and GB6 surveys. 

\begin{enumerate}

\item We compare the CIG radiocontinuum power at 1.4 GHz with the one of the 
UGC-SF sample  (Condon et al. \cite{condon02}) of field star-forming galaxies. The
sample is similar in distance and optically brighter. Taking into account the
upper-limits by using the survival analysis, the mean of $\log(P)$ for the CIG sample (20.11)
appears significantly lower than for the UGC-SF sample (21.61).

\item Radio-to-optical flux ratios ($R$) show a median value $\sim 4.5$,
confirming that radio emission in our sample is largely due to star
formation. Comparison with a sample of emission line galaxies (van
Duyne et al. 2004) reveals striking differences, with most $R$ values
between 10--100, while most of our sample lies in an order of
magnitude lower range, between $R=1$--10, similar to the one of the
UGC-SF sample.

\item The 1.4 GHz detection fraction for FIRST is much lower
than for NVSS even though FIRST is a more sensitive survey. This is
because the spatial frequency attenuation in FIRST makes it
sensitive only to compact, largely nuclear, emission in our sample
galaxies.  An isolated sample like AMIGA should be dominated by disk
star formation which FIRST cannot see. Thus NVSS, which is sensitive
to this emission, detects far more galaxies, and we estimate that
disk emission dominates the radio continuum signature of $\sim$80\%
of our largely spiral sample. Nuclear emission is thought to be
enhanced by interactions that dissipate angular momentum leading to
gas infall. Dissipative effects are likely near minimum in our
sample. This can be contrasted to surveys of compact groups (Menon
\cite{menon95}) where most radio detections involve compact nuclear emission
and where quasi-continuous tidal perturbations should efficiently
channel any unstripped gas into the centers of the galaxies.

\item Isolated early-type (E/S0) galaxies are one of the most
intriguing, albeit small ($\sim$14\%), parts of the AMIGA sample.
They show a similar detection fraction to that of the spirals which
already marks them as unusual. If they are true early-types then it
is difficult to ascribe their radio emission to star formation.
Either many of them are misclassified spirals, or isolated
early-types are somehow slightly hyperactive (unusual star
formation?, sites of minor mergers?). The 
number of radio detection of S0 galaxies is similar to the
ellipticals and show comparable radio luminosities. Since S0 show less
association with active nuclei these detections may be telling us that: 1)
some S0 could be  misclassified ellipticals and 2) others could be misclassified
spirals.

\item The RLF for our sample  was derived using the radio-optical
bivariate function. It also suggests that radio emission in our
isolated sample is star formation dominated. It shows a strong drop
at $P_{1.4{\rm{GHz}}}\sim 10^{23}$ W\,Hz$^{-1}$, with little contribution
from nuclear activity (see also Sabater et al. \cite{sabater08}). 

\item The RLF allowed us to compute the total radio power density
at 1.4 GHz for our sample. We found: $U_{1.4{\rm{GHz}}} \sim
4.4\times10^{18}$ W\,Hz$^{-1}$\,Mpc$^{-3}$, compared to $ 15.6\times
10^{18}$ for the IRAS-2 Jy sample, $ 17.1\times 10^{18}$ for the
UCG-SF sample, $ 41.9\times 10^{18}$ for the UGC-AGN sample, and $
74.8\times 10^{18}$ for the 2DFGRS sample. AMIGA gives at most 30\%
of the contribution of other star-forming samples and 10\% of
AGN-dominated samples.

\end{enumerate}

AMIGA provides a unique reference sample for studies that aim to
quantify the effects of environment. This is true whether one
studies interacting samples (e.g. pairs, compact groups) dominated
by one-on-one encounters or simply galaxies in environments with
different average surface densities.

\begin{acknowledgements}
We thank the anonymous referee for a careful reading and very detailed report which helped to improve this paper significantly.
SL, LVM, JSM, DE, UL, SV, GB and EG are partially supported by DGI Grant
AYA 2005-07516-C02 and Junta de Andaluc\'{\i}a (Spain). SL was partially supported
by an Averroes  Fellowship contract from the Junta de Andaluc\'ia. UL is supported by
a Ramon y Cajal fellowship contract and  acknowledges support
from the DGI grant  ESP2003-00915. GB is supported at the IAA/CSIC by an I3P
contract (I3P-PC2005-F) funded by the European Social Fund. 
This research has made use of the NASA/IPAC Extragalactic Database (NED)
which is operated by the Jet Propulsion Laboratory, California Institute of Technology, under contract with the National Aeronautics and Space Administration.
\end{acknowledgements}


\begin{thebibliography}{}
\bibitem[1973]{allen73}  Allen, R. J., Ekers, R. D., Burke, B. F., \& Miley, G. K. 1973, Nature, 241, 260
\bibitem[1995]{becker95}  Becker, R. H., White, R. L., \& Helfand, D. J. 1995, ApJ, 450, 559
\bibitem[1993]{benn93}  Benn, C. R., Rowan-Robinson, M., McMahon, R. G., Broadhurst,
  T. J., \& Lawrence, A. 1993, MNRAS, 263, 98
\bibitem[1996]{bertin96}  Bertin, E., \& Arnouts, S. 1996, A\&AS, 117, 393
\bibitem[1980]{condon80}  Condon, J. J. 1980, ApJ, 242, 894
\bibitem[1982]{condon82}  Condon, J. J., Condon, M. A., Gisler, G., \& Puschell, J. J. 1982, ApJ, 252, 102
\bibitem[1984]{condon84}  Condon, J. J. 1984, ApJ, 284, 44
\bibitem[1989]{condon89}  Condon, J. J. 1989, ApJ, 338, 13
\bibitem[1998]{condon98}  Condon, J. J., Cotton, W. D., Greisen, E. W., et al. 
1998, AJ, 115, 1693
\bibitem[2002]{condon02}  Condon, J. J., Cotton, W. D., \& Broderick, J. J. 2002, AJ, 124, 675
\bibitem[2005]{domingue05}  Domingue, D. L., Sulentic, J. W., \& Durbala, A. 2005, AJ, 129, 2579
\bibitem[1985]{feigelson85}  Feigelson, E. D., \& Nelson, P. I. 1985, ApJ, 293, 192
\bibitem[1996]{gregory96}  Gregory, P. C., Scott, W. K., Douglas, K., \& Condon, J. J. 1996, ApJS, 103, 427
\bibitem[1980]{hummel80}  Hummel, E. 1980, A\&A, 89, L1
\bibitem[1981]{hummel81}  Hummel, E. 1981, A\&A, 93, 93
\bibitem[1973]{karachentseva73}  Karachentseva, V. E. 1973, SoSAO, 8, 3
\bibitem[1987]{kennicutt87} Kennicutt, R. C. Jr., Roettiger, K. A., Keel, W. C., van der Hulst,
 J. M., \& Hummel, E.  1987, AJ, 93, 1011
\bibitem[1992]{lavalley92}  Lavalley, M., Isobe, T., \& Feigelson, E. 1992, ADASS I, 
ASP Conf. Series, Vol. 25, D. M. Worrall, C. Biemesderfer, and J. Barnes, eds., p. 245.
\bibitem[2003]{leon03}  Leon, S., \& Verdes-Montenegro, L. 2003, A\&A, 411, 391
\bibitem[2007]{lisenfeld07}  Lisenfeld, U., Verdes-Montenegro, L., Sulentic, J., et al.
2007, A\&A, 462, 507
\bibitem[1992]{menon92}  Menon, T. K. 1992, MNRAS, 255, 41
\bibitem[1995]{menon95}  Menon, T. K. 1995, MNRAS, 274, 845
\bibitem[1999]{menon99}  Menon, T. K. 1999, Ap\&SS, 269, 435
\bibitem[1974]{press74}  Press, W. H., \& Schechter, P. 1974, ApJ, 187, 425
\bibitem[1997]{rengelink97}  Rengelink, R. B., Tang, Y., de Bruyn, A. G., et al. 
1997, A\&AS, 124, 259
\bibitem[2008]{sabater08}  Sabater, J., Leon, S., Verdes-Montenegro, L. et al. 2008, A\&A in press
\bibitem[2002]{sadler02}  Sadler, E. M., Jackson, C. A., Cannon, R. D., et al. 2002, MNRAS, 329, 227
\bibitem[2002]{serjeant02}  Serjeant, S., Gruppioni, C., \& Oliver, S. 2002, MNRAS, 330, 621
\bibitem[1968]{schmidt68}  Schmidt, M. 1968, ApJ, 151, 393
\bibitem[1978]{stockeetal78}  Stocke, J. T., Tifft, W. G., \& Kaftan-Kassim, M. A. 1978, AJ, 83, 322
\bibitem[1978]{stocke78}  Stocke, J. T. 1978, AJ, 83, 348
\bibitem[1976a]{sulentic76a}  Sulentic, J. W. 1976a, AJ, 81, 582
\bibitem[1976b]{sulentic76b}  Sulentic, J. W. 1976b, ApJS, 32, 171
\bibitem[2003]{sulentic03}  Sulentic, J. W., Zamfir, S., Marziani, P., et al. 
2003, ApJ, 597, L17      
\bibitem[2006]{sulentic06}  Sulentic, J. W., Verdes-Montenegro, L., Bergond, G., et al.
 2006, A\&A, 449, 937
\bibitem[1968]{tovmassian68} Tovmassian, H. M. 1968, Australian Journal  of Physics, 21, 193
\bibitem[2004]{vanduyne04}  Van Duyne, J., Beckerman, E., Salzer, J. J., et al.
2004, AJ, 127, 1959
\bibitem[2005]{verdes05}  Verdes-Montenegro, L., Sulentic, J., Lisenfeld, U., et al.
 2005, A\&A, 436, 443
\bibitem[2007a]{verley07a}  Verley, S., Odewahn, S. C., Verdes-Montenegro, L., et al. 
2007a, A\&A, 470, 505
\bibitem[2007b]{verley07b}  Verley, S., Leon, S., Verdes-Montenegro, L., et al. 
2007b, A\&A, 472, 121
\bibitem[1974]{wright74} Wright, A. E. 1974, MNRAS, 167, 251
\bibitem[1991]{xu91}  Xu, C., \& Sulentic, J. W. 1991, ApJ, 374, 407
\bibitem[1989]{young89} Young, J. S., \& Knezek, P. M. 1989, ApJ, 347, L55 
\bibitem[2001]{yun01}  Yun, M. S., Reddy, N. A., \& Condon, J. J. 2001, ApJ, 554, 803
\end{thebibliography}
\end{document}